\documentclass[12pt,psf,epsf]{JHEP3}

\usepackage{graphicx}
\usepackage{graphics}

\usepackage{epsfig,bm}

\usepackage[small]{caption}

\usepackage{amsfonts,amssymb,amsmath,latexsym}

\newcommand{\be}{\begin{equation}}
\newcommand{\ee}{\end{equation}}
\newcommand{\bea}{\begin{eqnarray}}
\newcommand{\eea}{\end{eqnarray}}
\newcommand{\pd}{\partial}

\title{Cosmological Daemon}
\author{I.Ya. Aref'eva \\
\it Steklov Mathematical Institute, Russian Academy of Sciences,\\
Gubkina str. 8, 119991, Moscow, Russia \\
arefeva@mi.ras.ru}
\author{I.V. Volovich\\
\it Steklov Mathematical Institute, Russian Academy of Sciences,\\
Gubkina str. 8, 119991, Moscow, Russia \\
volovich@mi.ras.ru}
\abstract{Classical versions of the Big Bang cosmological models of the
universe contain a singularity at the start of time, hence  the time variable
in the field equations should run over a half-line.
Nonlocal string field theory equations with infinite number of derivatives
   are considered and
 an important difference between  nonlocal operators on the whole real line
and on a half-line is pointed out.
 We use the heat equation method and show that on the half-line in addition to the usual initial data a new arbitrary function (external source) occurs
that we call  the  daemon function. The daemon function  governs the
evolution of the universe similar to Maxwell`s demon in
thermodynamics. The universe and multiverse are open systems
interacting with the daemon environment. In the simplest case the
nonlocal scalar field reduces to the usual local scalar field
coupled with an external source which is discussed in the stochastic
approach to inflation. The daemon source can help to get the chaotic
inflation scenario with a small scalar field. }


\begin{document}

\newpage

\section{Introduction}
There are many considerations of various types  of boundary conditions in classical
and quantum cosmology \cite{Linde,Mukchanov,Weinberg,Rubakov}.
Classical versions of the Big Bang cosmological models of the universe
contain a singularity at the beginning of time, hence  the time variable
in the field equations should run over a half-line
\footnote{Rigorous mathematical definition of singularities in General Relativity
requires some boundary constructions, points at infinity and so
on \cite{Hawking-Ellis}.
Here we restrict ourself with a simple approach by considering the time
variable $t$ running over the half-line $[0,\infty)$ with regular boundary
conditions at $t=0$. }.

In this paper we consider nonlocal string field theory
equations with infinite number of derivatives
    and point out
 an important difference between  nonlocal operators on the whole real line
and on a half-line.  We use the heat equation
method and show that on the half-line in addition to the usual
initial data a new arbitrary function (external source) occurs
that we call  the  daemon function \footnote{According to Plato, daemons
are good or benevolent "supernatural beings between mortals and
gods, such as inferior divinities", and differ from the
Judeo-Christian usage of demon that is a malignant spirit. Socrates'
daimon is analogous to the guardian angel.}.
The daemon function  governs the evolution of the universe
similar to Maxwell`s demon in statistical physics which was created
to show that the second law of thermodynamics has only a statistical
certainty.  In the present approach the universe and even the Multiverse
are open systems interacting with the daemon environment.

In the simplest case the nonlocal
scalar field reduces to the usual local scalar field coupled
with an external source.
The daemon source can help to get  the chaotic inflation scenario
with a small  scalar field.

We shall consider the following string field theory effective action
\cite{IA04} describing the nonlocal scalar field coupled with
gravity:
\be
\label{NLCM} S=\int
d^4x\sqrt{-g}\left\{\frac{m_p^2}{2}R+ \Phi F(\Box)\Phi-V(\Phi)
\right\}, \ee where  $F(\Box)$ is a function of the D'Alembert
operator,
$\square=\frac1{\sqrt{-g}}\pd_{\mu}\sqrt{-g}g^{\mu\nu}\pd_{\nu}$,
$g_{\mu\nu}$ is the metric, $R$ is the scalar curvature, $m_p$
 is  a rescaled   Planck mass,
 $\Phi$ is a  scalar field and $V(\Phi)$ is a potential.
 Cosmological applications of various forms of such
 action have been considered
in \cite{IA04}-\cite{IA:2011}.

Nonlocal operators that appear in string field theory
\cite{SFT-review} and in $p$-adic strings
\cite{p-adic-string,VVZ,padic-review}\footnote{$p$-adic cosmology
is considered in \cite{ADFV}.} usually
include the exponential function of the D'Alembert operator,
$e^{-\Box}$.

There are two important properties of the function $F(z)$ to be
discussed. First, if the equation $F(z)=0$ has a number of roots
$\{z=m^2_n\}$ then our theory with the nonlocal field $\Phi$ is
equivalent  to the theory with the
corresponding number of local fields $\phi_n$ \cite{AI-KAS:2006,AI-IV-NEC,AJV:2007a,AI-IV-Riem,AJV:2007b,MN:2008}.

Second, and it is the subject of this paper, if  the function
$F(\Box)$ includes the exponential function of the D`Alembertian,
$F(\Box)=e^{-\Box}+...$, then, in the case when we work on the
half-line $t>0$,
 an additional freedom in  the boundary conditions at the point $t=0$ occurs,
besides the usual initial data. In the homogenous case we have to
define the action of the operator $ e^{\partial_t ^2}$ on the
function $\varphi(t)$. To this end, one can
 use an auxiliary function of two variables $\Psi(\tau,t)=
 e^{\tau\partial_t ^2}\varphi(t)$
satisfying  the  heat  equation
\be \label{HE-1} (\partial
_\tau-\partial_t ^2)\Psi(t,\tau)=0,\ee with the initial condition
\be \label{BC-1} \Psi(t,\tau)|_{\tau=0}=\varphi(t).
\ee Then the
action of $e^{\partial_t ^2}$ on the function $\varphi(t)$
 is defined by   the  solution to (\ref{HE-1})
with the initial data (\ref{BC-1}) as follows
\be \label{BC1}
e^{\partial_t ^2}\varphi(t)=\Psi(t,\tau)|_{\tau=1}.
\ee
This
prescription works if  the time variable $t$ runs over  the whole
real line. The heat equation (diffusion equation) on the whole line
is an efficient  method  to study the rolling tachyon solutions
in flat space \cite{MZ,YaV,AJK,YaVV,VSV,Cal_Nard_2008} as well as in the FRW cosmology
\cite{Joukovskaya:2007nq,MN:2008,LJ:2008,Cal_Nard_JHEP2010}.  It is important to
note that on the half-line $t>0$ we have to add to the initial data
(\ref{BC-1}) also the boundary condition at $t=0$ with some function
$\mu$.

We consider the heat equation (\ref{HE-1}) on the half-line $t>0$
and $\tau>0$ with the initial data (\ref{BC-1}) and the following
boundary condition at $t=0$: \be \label{HE-bc-ic-1}
\Psi(0,\tau)=\mu(\tau). \ee This is the mixed initial-boundary value
Dirichlet's problem for the heat equation on the half-line. The
function $\mu(\tau)$ will be called the Dirichlet daemon function.
It describes the Dirichlet boundary conditions at $t=0$. On the
half-line $t>0$ we define the action of the operator
 $e^{\partial_t ^2}$ on the function $\varphi(t)$
  as $e^{\partial_t ^2}\varphi(t)= \Psi(t,\tau)|_{\tau=1}$,
  where $ \Psi(t,\tau)$ is the solution of the initial-boundary
  value problem (\ref{HE-1}),(\ref{BC-1}), (\ref{HE-bc-ic-1}). Now the result
  $e^{\partial_t ^2}\varphi(t)$  depends on the function $\mu$, see
  Sect.3.

For the nonlocal action  (\ref{NLCM}) in addition to the local
fields $\phi_n$ the new arbitrary function $\mu$ appears  which is
an external source. The simplest version of the local action with
an external source has the form
\begin{equation}
S=\int d^4x\sqrt{-g}\left\{\frac{M_P^2}{2}R+ \frac{1}{2}
\phi\Box\phi-U(\phi)+J\phi \right\}, \label{L-l-inf-pairs}
\end{equation}
where $J=J(x)$ is the external source and $U(\phi)$ is the
potential. We shall discuss the inflation  scenario for this action
in the conclusion.

Note that if the source $J(x)$ is a white noise then the  equations
of motion for the local action (\ref{L-l-inf-pairs}) are reduced to
the Langevin equations in the stochastic approach to inflation
introduced by Starobinsky \cite{Starob}.

In mathematical literature equations with infinite number of
derivatives have been considered in \cite{YaVV}-\cite{Gorka} (see
also \cite{davis,Carmichael,Carleson,Hormander}),
 but, as we are aware of,
the appearance  of the additional boundary function $\mu$ on the half-line
for the exponential operator was not discussed. Integral equations at the half-line
are investigated in \cite{Khach} (and ref's therein).

 The operator $e^{\partial_t ^2}$ on the half axis by using the Laplace
transform has been considered in the stimulating
paper by  Barnaby and  Kamran \cite{BK:2007}. Note that the Mellin transform has been
also used in the investigation of the quantum Riemann function
 $\zeta (\Box)$ \cite{AI-IV-Riem}.
In the present paper we clarify the relation between these two approaches:
the Laplace transform
 and the heat equation.
We show in Sect.5 that the Laplace transform method is a special case of the more
general heat equation method and it does not include an arbitrary
function $\mu(\tau)$
at the boundary $t=0$. The Laplace transform method corresponds to a special choice
of the daemon function $\mu(\tau)$ that is uniquely  defined by the function
$\varphi(t)$.

The paper is organized as follows. In Sect.2 we remind the  form
of nonlocal string field theory and p-adic string actions. In Sect.3 we present the
definition of the operator $e^{-\Box}$ via
the heat (diffusion) equation and stress the difference between the diffusion
equation on the whole real line  and on the half-line.
 In Sect.4 we show that on the half-line in addition to the usual initial data
 a new arbitrary function (external source) occurs
that we call  the  daemon function. Depending of the type of the boundary conditions,
the Dirichlet or Neumann, we deal with different definitions
 of the operator $e^{\pd _t^2}$.
 In Sect.5 we show that interpretation  of the operator $e^{\pd_ t^2}$
 as a symbol via Laplace transform corresponds to  the Dirichlet  boundary
 condition with a special source. In Sect.6 we present solutions to a linearized
 nonlocal equation corresponding to the Dirichlet and Neumann boundary conditions.
 In Sects.7 and 8 we describe  approximate solutions to  nonlinear
 equations. In Sect. 9 we give a generalization of
  the previous results to the case of the FRW metric.
In Sect.10 we sketch  possible applications of the external daemon source
for inflation scenario, namely we discuss
the chaotic and stochastic approaches to inflation in the presence
of an external source.

\section{Nonlocal String Field and p-adic String Actions}
The level truncated string field action
for  the tachyon leaving on
 the 3-brane and interacting in a minimal way with gravity
has the form \cite{IA04}
\be
\label{SFT-t}
S=\int
d^4x\sqrt{-g}\left\{M_p^2\frac{R}{2}+\frac{1}{\alpha ^{\prime 2}g^2_o}
\left(\frac{\alpha ^{\prime}}{2}
\Phi (\Box +\frac{1}{\alpha ^{\prime}})e^{-\kappa \alpha ^\prime\Box}\Phi-
V(\Phi)\right)\right\}.
\ee
Here $\alpha ^\prime$ is a square of the characteristic length  of the string
and $1/\alpha ^\prime$ defines the mass of tachyon,
 $\Phi$ is a dimensionless tachyon field, $V(\Phi)$ is the tachyon potential,
 $V(\Phi)=\frac{1}{3}\Phi^3$ for the bosonic string,   $V(\Phi)=\frac{1}{4}\Phi^4$ for the fermionic
  string, $g^2_0$ is the
 dimensionless open string coupling constant, $\kappa$ is a numerical parameter that is fixed by the string field theory,
$\kappa=2\ln(4/3\sqrt{3})$.

The $p$-adic open string tachyon action  has the form
\be
\label{p-adic}
S=\int
d^4x\sqrt{-g}\left\{M_p^2\frac{R}{2}-\frac{1}{\alpha ^{\prime 2}g_p^2}
\left(\Phi
e^{-\kappa_p\, \alpha ^\prime\Box}\Phi-
V(\Phi)\right)\right\}.
\ee
Here we also assume that the characteristic scale of the theory is $\alpha^\prime$,
$g_p$ is a coupling constant, $V(\Phi)=\frac{1}{p+1}\Phi^{p+1}$.

Using dimensionless coordinates, $x/\sqrt{\alpha^\prime}$
and performing recsaling one rewrites these actions in the form
 \be
\label{string}
S=\int
d^4x\sqrt{-g}\left\{m_p^2\frac{R}{2}+\frac{1}{\gamma}\left(\frac1{2}\Phi(\xi^2\Box+1)
 e^{-\Box }\Phi-
V(\Phi)\right)\right\}.
\ee
Here $ m_p^2=\frac{M_p^2}{M_s^2} $,
$\gamma=g_o^2$ for action (\ref{SFT-t}) and  $\gamma=-g_p^2<0$ in
the p-adic inflation models \cite{Lidsey07,Barnaby:07}. Note that in
a naive local approximation ($e^{-\Box }\approx 1-\Box $)
the case $\xi=0$, $V=0$ with  $\gamma
<0$ corresponds to a massive field and with $\gamma>0$ corresponds to a
ghost field with a negative mass term.

Therefore, string field theory (SFT) and $p$-adic string lead to (\ref{NLCM}) with the following form of
$F(\Box)$, \cite{IA04}
\be F(\Box)=(\xi^2\Box+1 )e^{-\alpha \Box}-\beta.\ee

As it has been mentioned in Introduction there are two possibilities
 to understand $e^{-\frac14\Box }$
as an operator on the  whole real line and only on the semi-line.

It has been shown,
 \cite{AI-IV-NEC,AJV:2007a,AI-IV-Riem} that   the
action (\ref{NLCM}) with the nonlocal field $\Phi$ on the whole axis
is equivalent, modulo total derivatives, to the following
action with an infinite  number of local fields
$\psi_n$
\begin{equation}
S=\int d^4x\sqrt{-g}\left\{\frac{M_P^2}{2}R+
\frac12 \sum_n
\left[\eta_n \psi_n e^{f(\Box)}(\Box-\alpha_n^2)\psi_n +
c.c.\right]-U(\{\psi_n\})\right], \label{l-inf-pairs}
\end{equation}
where $\eta_n, \alpha_n$ are constants and $f$ is a function.
This is a generalization of the similar
flat result
\cite{PaisU:1950}.
To get this representation the following Weierstrass product
for the function $F(z)$ has been used: $F(z)=e^{f(z)}\prod_n (z-\alpha_n^2)$
(more general Weierstrass representations are discussed in \cite{AJV:2007a,AI-IV-Riem}).

In particular, for $\xi=0$
\be F(\Box)=e^{-\alpha \Box}-\beta,\ee
where $\alpha,\beta$
are constants, there is the  following Weierstrass representation,
see \cite{AI-IV-NEC}
\be \label{Non8}
F(\Box)=e^{-\alpha\Box}-\beta=-\beta^{1/2}e^{-\alpha\Box/2}
(\alpha\Box+\log\beta)\prod_{j=1}^{\infty}
(1+\frac{(\alpha\Box+\log\beta)^2}{4\pi^2 j^2}).
 \ee

\section{Nonlocal operator $e^{\partial_t ^2}$ and  heat equation }

The D'Alembert operator in Minkowski space has the form
\be
\Box=-\partial^2_t+ \partial_{x_i}\partial_{x_i}.
\ee
 In the homogenous case we have to give
a mathematical meaning to the operator
$
e^{\partial_t ^2}.$
To this end many authors have used the Fock auxiliary parameter method
\cite{Fock,Feynman}. This method is known also as the heat equation method,
or the diffusion equation method.
This method
 uses an auxiliary function of two variables $\Psi(\tau,t)=
 e^{\tau\partial_t ^2}\varphi(t)$
satisfying  the  heat (diffusion) equation \cite{YaVV,VSV,MZ,Joukovskaya:2007nq,Barnaby:2008pt,Cal_Nard_2009,Cal_Nard_JHEP2010}
\be
\label{HE}
(\partial _\tau-\partial_t ^2)\Psi(t,\tau)=0,\ee
and the initial condition
\be
\label{BC}
\Psi(t,\tau)|_{\tau=0}=\varphi(t).
\ee
Note that originally $t$ is temporal variable but  here in the heat equation  the variable $t$
plays the role of the spatial variable and $\tau$ is "time". Then the action of $e^{\partial_t ^2}$ on a function $\varphi(t)$
 is defined by means of the  solution to (\ref{HE})
with the boundary condition (\ref{BC}) as follows
\be
\label{BC1}
e^{\partial_t ^2}\varphi(t)\equiv \Psi(t,\tau)|_{\tau=1}.\ee

Now it is important  to distinguish two  different cases:
\begin{itemize}
\item i)  whole real line $-\infty <t <\infty$,
\item ii) half-line $0<t< \infty$.
\end{itemize}

In quantum field theory in Minkowski space-time the time variable $t$ runs from $-\infty$ to
$\infty$. If we are interested in application of SFT  equations, or p-adic string equations
 to cosmology
 \cite{IA04,Barnaby:06,Lidsey07} it is important to stress that there is a reason  to restrict the time variable to
the half axis only,
$t>0$, since in the Friedman cosmology there is a cosmological singularity
at $t=0$ \cite{Hawking-Ellis}\footnote{Bouncing solutions can be extended to the
 whole real line \cite{AJ:2005}}.

\section{Dirichlet's Daemon and Neumann's Daemon}

In the case of the whole real axis under suitable assumptions there is a unique
solution of the problem (\ref{HE}), (\ref{BC}). However,
in the case of the half-line one has to add an extra boundary condition
at $t=0$, see for example \cite{VSV-textbook}. Therefore, we shall consider in this note
the following mixed initial-boundary value problem:
\be
\label{HE-bc-ic}
\begin{cases}\frac{\partial}{\partial \tau}\Psi_D(t,\tau)=
\frac{\partial^2}{\partial t^2}\Psi_{D}(t,\tau),\,\,\,\,\,t>0,\,\,\,\,\tau>0,\\
\Psi_{D}(t,0)=\varphi(t),\\
\Psi_{D}(0,\tau)=\mu(\tau).
\end{cases}\ee
This is the Dirichlet's problem for the heat equation on the half-line.
The function $\mu(\tau)$ will be called the Dirichlet daemon function.
It describes the Dirichlet boundary conditions at $t=0$.
Our goal is to evaluate the contribution of the function
$\mu(\tau)$ in the definition of the operator $e^{\partial_t ^2}$, $\,t>0$
to cosmological applications. In the heat equation the physical meaning of the
boundary function $\mu$ is that it describes of how the
temperature at the boundary  depends on time. In our case the role of time
plays an auxiliary parameter $\tau$.

We define the action of the operator $e^{\tau\partial_t ^2}$
on the half axis as follows
\be
\label{Bcr}
e^{\tau\partial_t ^2}\varphi(t)= \Psi_{D}(t,\tau),\,\,\,\,\,t>0,\,\,\,\tau>0,\ee
where $\Psi_{D}(t,\tau)$ is a solution to (\ref{HE-bc-ic}).
To solve the Dirichlet problem for the heat equation on the half-line
one uses the odd extension of $\varphi$  to the whole real line.
It is known \cite{VSV-textbook} that the solution of  (\ref{HE-bc-ic})
is given by
\bea
\nonumber
\Psi_{D}(t,\tau)&=&\frac{1}{\sqrt{4\pi \tau}}\int _0^\infty \,\varphi(t^\prime)\,
[e^{-\frac{(t-t')^2}{4\tau}}-e^{-\frac{(t+t')^2}{4\tau}}]dt'\\
&+&\frac{t}{\sqrt{4\pi }}\int _0^\tau\,
\frac{\mu(\tau')}{(\tau-\tau')^{3/2}}e^{-\frac{t^2}{4(\tau-\tau')}}\,d\tau'
\label{TB-sol}.\eea

Our definition of the action of operator
$e^{\lambda\partial^2_t}_{D,\mu}$ with any function  $\mu$ can
be presented as
\be
\label{e-D-mu}
e^{\tau\partial^2_t}_{D,J}\varphi (t)=e^{\tau\partial^2_t}_{\,\,\,0}\varphi(t)
+J(t,\tau),\ee
where
\be
\label{TB-sol-r-1}
e^{\tau\partial^2_t}_{D,0}\varphi=
\frac{1}{2\sqrt{\tau\pi} } \int\limits_0^\infty
dt'\,[e^{-\frac{(t-t')^2}{4\tau}}-e^{-\frac{(t+t')^2}{4\tau}}] \,\varphi(t')\ee
and the source $J$ is given by
\be
J(t,\tau)=\frac{t}{\sqrt{4\pi }}\int _0^{\tau}\,
\frac{\mu(\tau')}{(\tau-\tau')^{3/2}}e^{-\frac{t^2}
{4(\tau
-\tau')}}\,d\tau'.
\label{TB-sol-r-2}
\ee
Here the subscript  $D$ means the Dirichlet and we
write the index $J$ to stress that the definition
depends on the choice of the daemon source $J$.

Consider also the Neumann problem for the heat
equation on the half-line $t>0$:

\be
\label{HE-bc-ic-N}
\begin{cases}\frac{\partial}{\partial \tau}\Psi_N(t,\tau)=
\frac{\partial^2}{\partial t^2}\Psi_N(t,\tau),\,\,\,\,\,t>0,\,\,\,\,\tau>0,\\
\Psi_N(t,0)=\varphi(t),\\
\frac{\partial}{\partial t}\Psi_N(t,\tau)|_{t=0}=\nu(\tau).
\end{cases}\ee
To solve the Neumann problem on the half-line one extends $\varphi$
to the whole line in such a way that the extension is even and gets the solution:
\bea
\nonumber
\Psi_N(t,\tau)&=&\frac{1}{\sqrt{4\pi \tau}}\int _0^\infty \,\varphi (t')\,
[e^{-\frac{(t-t')^2}{4\tau}}+e^{-\frac{(t+t')^2}{4\tau}}]dt'
\label{sol-n-der}\\
&-&\frac{1}{\sqrt{\pi }}\int _0^{\tau}\,
\frac{\nu(\tau')}{(\tau-\tau')^{1/2}}e^{-\frac{t^2}{4(\tau-\tau ')}}d\tau'.
\eea

We define the operator $e^{\tau\partial^2_t}$ on the half-line with
the Neumann boundary condition as \be
e^{\tau\partial^2_t}_{N,J}\varphi(t)=\Psi_N(t,\tau). \ee Here $N$
means the Neumann boundary condition and the source $J$ is given by
\be
\label{current-N}
J(t,\tau)=\frac{1}{\sqrt{\pi }}\int _0^{\tau}\,
\frac{\nu(\tau')}{(\tau-\tau')^{1/2}}e^{-\frac{t^2}{4(\tau-\tau
')}}d\tau'
\ee
with the daemon function $\nu$.

There are also other initial-boundary problems for the heat equation
on the half-line but we will not discuss them in this paper.

\section{Laplace Daemon }

In this section we show that the  definition of the operator
$e^{
\partial_t ^2}$ on the half-line based on the Laplace transform
is a special case of our definition (\ref{TB-sol}) based on the
solution of the
heat equation with mixed boundary conditions.

The Laplace transform method
 for the definition of the operator
$e^{
\partial_t ^2}$ on the half-line is used in \cite{BK:2007}, see Appendix 1 for more details.

We apply the general definition (\ref{ps-F}) to the case of exponential function and
obtain
the following definition
\be \label{pseudoapp-exp}
  e^{\tau\partial^2_t}_{\,\,\,\{L\}}\varphi(t) =
  \frac{1}{2\pi i}\int\limits_{c-i\infty}^{c+i\infty} d\!s\, e^{st}\left[ e^{\tau s^2}  \tilde{\varphi}(s)
  -   \sum_{k=1}^{\infty} \sum_{j=1}^{2k}
  \frac{\tau^ks^{2k-j}}{k!}\varphi^{(j-1)}(0)  \right].
  \ee
Here the subscript $L$ indicates that the definition is based on the Laplace transform.

Now we show that the definition (\ref{pseudoapp-exp}) based on the
Laplace transform is a particular  case of
the definition (\ref{TB-sol})
based on the mixed problem for the heat equation with the special  choice
of the function
$\mu$.

{\bf Proposition 1.} {\it The following relation takes place}

\be
\label{prop1}
e^{\tau\partial^2_t}_{\,\,\,\{L\}}\varphi(t)=e^{\tau
\partial_t ^2}_{D,J}\varphi(t),\ee
{\it  where} $J$ {\it is}
\be
J(t,\tau)=\frac{t}{4\sqrt{\pi }}\int\limits_0^\tau \,d\tau'
e^{-\frac14\frac{t^2}{(\tau-\tau')}}\, \frac{\phi
(\sqrt{\tau' })+
 \phi (-\sqrt{\tau' })}{(\tau-\tau')^{3/2}}
 \ee
{\it and }$\phi (x)$ {\it is defined by a map}
\be
\varphi (x)\,\, \to \,\,\,\phi (x)\ee
{\it by using the following rule. If the function }
$\varphi (x)$ {\it is given by
the series}
\be
\varphi (x)=\sum _{n=0}^\infty \varphi_n\frac{x^n}{n!}\ee
{\it then}
 \be
 \label{phi-varphi}
\phi (x)=\sum _{n=0}^\infty \varphi_n\frac{x^n}{[\frac{n}{2}]!}.\ee

{\it Formula (\ref{prop1}) can be written more explicitly},
\bea
\label{Act} e^{\tau\partial^2_t}_{\,\,\,\{L\}}\varphi(t)
&=&\nonumber\frac{1}{2\sqrt{\tau\pi} } \int\limits_0^\infty
dt'\,[e^{-\frac{(t-t')^2}{4\tau}}-e^{-\frac{(t+t')^2}{4\tau}}] \,\varphi(t')\\
\nonumber
\\&+&\frac{t}{4\sqrt{\pi }}\int\limits_0^\tau \,d\tau'
e^{-\frac14\frac{t^2}{(\tau-\tau')}}\, \frac{\phi
(\sqrt{\tau' })+
 \phi (-\sqrt{\tau' })}{(\tau-\tau')^{3/2}}.
\eea

This proposition shows that the definition of the operator
$e^{\tau\partial^2_t}_{\,\,\,L}$ which uses the Laplace
transform is a particular
 case of the definition of the operator $e^{\tau
\partial_t ^2}_{ \{\mu\}}$ which uses the solution of the heat equation
with the special daemon function $\mu$.

!!The proof of Proposition 1 is given in Appendix B.!!
!!So, one can say that Laplace daemon is subordinated to Dirichlet's daemon.!!

\section {Solutions to Linearized Nonlocal Equation on the Half-line}
\subsection{ Dirichlet's daemon without source}
Here we first consider the case of the Dirichlet's
daemon with  $j(\mu|t,\tau)= 0$.
In this case the equation of motion
\be\label{D0-m}
e^{\tau\partial^2_t}_{D,0}\Phi(t)=m^2\Phi(t), \,\,\,t>0,
\ee
where $m>0$ and $\tau >0$ are fixed parameters,
 has the form of the  following integral equation
\be
\frac{1}{2\sqrt{\tau\pi} } \int\limits_0^\infty
dt'\,[e^{-\frac{(t-t')^2}{4\tau}}-e^{-\frac{(t+t')^2}{4\tau}}] \,\Phi(t')
=m^2\Phi(t),\,\,\, t>0.
\ee
One can solve  equation (\ref{D0-m}) by making the odd extension of the function
$\Phi(t)$ to the whole real line. Then we get the equation
\be\label{DR}
e^{\tau\partial^2_t}\Phi(t)=m^2\Phi(t),\,\,\,\,t\in \mathbb{R}.
\ee
The odd solution of Eq. (\ref{DR})
is given by a linear combination
\be
\label{Phi-alpha}
\Phi_D(t)=\sum _n \left(\frac12\,B_n (e^{\alpha_nt}
-e^{-\alpha_nt})+\frac12\,B^*_n (e^{\alpha^*_nt}-e^{-\alpha^*_nt})\right),\ee
where $\alpha_n$ are solutions to the equation
\be
\label{m-eq}
e^{\tau \alpha^2_n}=m^2,\ee
\be
\label{alpha-n}
 \alpha_n=\pm\sqrt{\frac{\ln m^2+2\pi in}{\tau}}, n=0,\pm 1,\pm 2,...\ee
For the $\alpha_0$ mode,
\bea
\label{Phi-rs}
\Phi(t)=
B_0 \sinh \alpha_0t \eea
we have
\be\label{D0-1-a}
e^{\tau\partial^2_t}_{D,0}\sinh rt=e^{\tau r^2}\sinh rt, \,\,\,\,\,t>0,
\,\,\,\,\,e^{\tau r^2}=m^2.
\ee
Indeed,
the action of the operator $e^{\tau\partial^2_t}_{\,\,\,0}$
 on the function $e^{rt}$ is given by the formula
\bea
\label{act-exp}
e^{\tau\partial^2_t}_{\,D,0}e^{rt}&=&
\frac12e^{rt+r^2\tau}\left[1+\mbox{erf}\left(r\sqrt{\tau}+\frac{t}
{2\sqrt{\tau}}\right)\right]-
\frac12e^{-rt+r^2\tau}\left[1+\mbox{erf}\left(r\sqrt{\tau}-\frac{t}
{2\sqrt{\tau}}\right)\right]
\nonumber\eea
since
\bea
\nonumber
&\,&\frac{1}{2\sqrt{\tau\pi} } \int\limits_0^\infty
dt'\,[e^{-\frac{(t-t')^2}{4\tau}}-e^{-\frac{(t+t')^2}{4\tau}}]\,e^{\,rt'}\\
\label{int-ext}&=&\frac12e^{rt+r^2\tau}\left[1+\mbox{erf}\left(r\sqrt{\tau}+\frac{t}
{2\sqrt{\tau}}\right)\right]-
\frac12e^{-rt+r^2\tau}\left[1+\mbox{erf}\left(r\sqrt{\tau}-\frac{t}
{2\sqrt{\tau}}\right)\right]
\eea
Hence, we get (\ref{D0-1-a}).

 The energy and pressure on the solution have the form \cite{AJV:2007a}
\bea
E&=&-\frac{m^2\ln m^2}{2} { B}^2_0,\\
p&=&-\frac{m^2\ln m^2}{2}e ^{2\alpha_0^2}{ B}^2_0\cosh (2\alpha _0t),
\eea
where we set  $\tau=2$ and $g_p=1$. The energy is constant but the pressure depends on time.
We see that if $m^2>1$ the energy $E$ is negative, but $-(E+p)$
that defines the derivative of the Hubble parameter (see Sect. 7)
$\dot H$ is positive,
\be
\dot H=-\frac{1}{2m_p^2}(E+p)=
\frac{m^2}{2m_p^2}\frac{\ln m^2}{2}{ B}^2_0(1+\cosh (2\alpha _0t)).
\ee
Note that in the naive local approximation  to (\ref{p-adic}) we deal with a massive ghost field
and an appearance of a negative energy is not surprising. If we take in (\ref{p-adic})
$g_p=-1$ we see that the local approximation  corresponds to the usual massive field and in this case we get the positive energy.

We will see in the next subsection that in the case of the Neunmann daemon
we get the opposite signs.

\subsection{ Dirichlet's daemon with source}
Now we consider the nonlocal equation with the Dirichlet boundary conditions
and with an external  source
\be
e^{\tau\partial^2_t}_{\,D,J}\Phi=m^2\Phi. \label{j-m}
\ee
We are going to find a  solution to this equation for a special form of
the daemon source $J$.

The action of the operator
$e^{\tau\partial^2_t}_{D,0}$ on the constant $C$ is given by
\be
e^{\tau\partial^2_t}_{\,D,0}C=
C{\mbox{erf}}(\frac{1}{2}\frac{t}{\sqrt{\tau}}),\ee
where
$
{\mbox{erf}}(t)$ is
 the error function, (\ref{erf}).

Acting by $e^{\tau\partial^2_t}_{\,D,0}$ to $\Phi_0+B_0 \sinh(rt)$
we get
\be
e^{\tau\partial^2_t}_{\,D,0}(\Phi_0+B_0 \sinh(rt))=
\Phi_0{\mbox{erf}}(\frac{1}{2}\frac{t}{\sqrt{\tau}})+B_0e^{\tau r^2} \sinh(rt)\label{an-B0B1}\ee
This relation we can interpret as
\be
e^{\tau\partial^2_t}_{\,D,0}\Phi=-J(\tau,t)+m^2\Phi \label{j-m-D}
\ee
for $J(\tau,t)$ and $\Phi(t)$ in the form
\bea
\label{j01}
J(\tau,t)&=&m^2\Phi_0-\Phi_0{\mbox{erf}}(\frac{1}{2}\frac{t}{\sqrt{\tau}}),
\\
\Phi(t)&=&\Phi_0+B_0 \sinh(rt),\label{an-B0B1}
\eea
and
\be
r^2=\frac{\ln m^2}{\tau}.
\ee
According to (\ref{e-D-mu})  we rewrite (\ref{j-m-D}) as (\ref{j-m}).

\subsection{ Neumann's daemon without source}
Here we consider the case of the  Neumann daemon with  $J= 0$.
In this case solutions to the linearized equation of motion
\be
e^{\tau\partial^2_t}_{N,0}\Phi=m^2\Phi. \label{eom-d-0}
\ee
is given by a linear combination
\be
\Phi(t)=\sum _n \left(\frac12\,C_n (e^{\alpha_nt}+e^{-\alpha_nt})+\frac12\,C^*_n (e^{\alpha^*_nt}+e^{-\alpha^*_nt})\right)\ee
where  $\alpha_n$ are given by equation (\ref{alpha-n}).
For the $\alpha_0$ mode
\be\Phi(t)=
C_0 \cosh \alpha_0t
\ee
we have
\be\label{D0-1}
e^{\tau\partial^2_t}_{D,0}\cosh \alpha_0t=e^{\tau \alpha_0^2}\cosh \alpha_0t,
\,\,\,\,\,t>0.
\ee
The energy and pressure on this one mode approximation  are \cite{AJV:2007a}
\bea
E_{N}&=&
\frac{m^2\ln m^2}{2} C^2_0,\\
p_{N}&=&-\frac{m^2\ln m^2}{2}C^2_0\cosh (2\alpha _0t).
\eea
We see that if $m^2>1$ the energy $E$ is positive as well as   $-(E+p)$, and
consequently the derivative of the Hubble parameter
$\dot H$ is positive,
\be
\dot H=-\frac{1}{2m_p^2}(E+p)=
\frac{m^2}{2m_p^2}\frac{\ln m^2}{2} C^2_0(-1+\cosh (2\alpha _0t)).
\ee
This result is rather surprising, since we see that the energy is positive in the case
corresponding to  the case of the naive massive ghost mode.

\section{Nonlinear Tachyon Equation on the Whole Line}

It is instructive to compare the approximate solution obtained in
the next section for the nonlocal equation on the half-line with the
solution for the nonlinear tachyon equation on the whole line:

\be
\label{nonlin-nonloc-tachyon}
e^{\tau \partial ^2}(\xi^2\partial ^2-\mu ^2)\Phi(t)=-\epsilon
\Phi^3(t), \,\,\,\,\,\,\,\,\tau, \mu, \epsilon >0. \ee
This equation on the whole line can be written as an integral equation

\be
\frac{1}{2\sqrt{\tau\pi} } \int\limits_{-\infty}^\infty
dt'\,\left( \frac{\xi^2(( t-t' ) ^2-2\tau)}{4\tau^2}-\mu ^2\right)\,e^{-\frac{(t-t')^2}{4\tau}} \,\Phi(t')
=-\epsilon\Phi^3(t).
\ee
This equation for $\xi=0$ an an odd function  $\Phi(-t)=-\Phi(t)$  has the form
\be
\frac{1}{2\sqrt{\tau\pi} } \int\limits_{0}^\infty
dt'\,\left( \,e^{-\frac{(t-t')^2}{4\tau}}-e^{-\frac{(t+t')^2}{4\tau}}\right) \,\Phi(t')
=\epsilon\Phi^3(t).
\ee
It is known \cite{YaVV} that this equation has a solution interpolating
 between two vacua $\Phi=\pm \frac{\mu}{\sqrt{\epsilon}}$.

It is also known \cite{YaV} that for $\xi>\xi _{cr}$  equation
(\ref{nonlin-nonloc-tachyon}) has oscillation solutions with period $T_\xi$ and within
one period a good approximation for these solutions is

\be
\label{nl-series}
\Phi(t)=a\sinh\left(\Omega \,t\,\right)-
\frac{\epsilon a^3}{32\mu^2}\,e^{-9\lambda \Omega^2}\,\sinh\left(3\Omega\, t\,\right)+...
\ee
where $\Omega $ is a root of
\be
\label{Omega-lambda}
\Omega^2-\frac{\mu^2}{\xi^2}-\frac34\frac{\epsilon}{\xi^2} a^2\,
e^{-\lambda \Omega^2}=0.\ee
Approximation (\ref{nl-series}) is valid  for small $\lambda$ and small $t$, $t<T_\xi/4$.
Solutions to (\ref{Omega-lambda})
are given by the Lambert  W function satisfying a relation $W(x)e^{W(x)} = x$,
\be
\label{Omega-lambda-m}
\Omega=\pm\,\sqrt{\frac{\mu^2}{\xi^2}+\frac{1}{\lambda}{\rm {W}}\left(\frac{3\lambda\epsilon a^2}{4\xi^2} \, e^{-\frac{\mu^2}{\xi^2}\lambda}\right)}\ee
Expanding the right hand site of (\ref{Omega-lambda-m}) on $\lambda$ we get
\be
\Omega=\pm\left(\sqrt{\frac{\mu^2}{\xi^2}+\frac34 \frac{\epsilon}{\xi^2} a^2}-\frac38\frac{\epsilon}{\xi^2} a^2\lambda\sqrt{\frac{\mu^2}{\xi^2}+\frac34 \frac{\epsilon}{\xi^2} a^2}
+{\cal O}(\lambda^2)\right)
\ee
Note, that if $\lambda=0$ one gets an approximate solution to the local equation
\be
\label{nonlin-tachyon} (\xi^2\partial ^2-\mu ^2)q(t)=-\epsilon
q^3(t)
\ee

\be
\label{nonlin-tachyon-APV} q(t)=A\sinh\left(\frac{\mu^2}{\xi^2}(t-\frac
{3A^2}{8\mu^2}\epsilon t)\right)+ \frac{\epsilon
A^3}{32\mu^2}\sinh\left(\frac{3\mu^2}{\xi^2}(t-\frac {3A^2}{8\mu^2}\epsilon
t)\right)+... \ee
Note an appearance of the hyperbolic functions
instead of the trigonometric functions in the Bogolyubov-Krylov
averaging  method for the nonlinear oscillations \cite{BK}.

\section{Nonlinear Nonlocal Equations on  the Half-line}
\subsection{Dirichlet's case}
In this section we consider a nonlinear nonlocal equation on  the half-line
of the following form
\be
e^{\tau\partial^2_t}_{\,D,J}\Phi=V^\prime (\Phi(t)), \label{nl-j-m}
\ee
where $e^{\tau\partial^2_t}_{\,D,J}$ is given by (\ref{e-D-mu}).
By taking $t=0$ in (\ref{nl-j-m}) we obtain
\be\label{DG}
\mu(\tau)=V^\prime (\Phi(0)).
\ee
Equation (\ref{DG}) shows that the initial value of the field $\Phi (0)$ is governed by
the Dirichlet daemon function $\mu$.

Let us find an approximate  solution to (\ref{nl-j-m}) for
$V(\Phi)=\frac{g}{3}\Phi^3$. We consider the equation
\be
\label{eq-3-j}
e^{\tau\partial^2_t}_{\,D,0}\Phi=-J+g\Phi^2\ee
for $J$ in the form
\be
\label{j01m}
J(\tau,t)=\mu_0+\mu_1{\mbox{erf}}(\frac{1}{2}\frac{t}{\sqrt{\tau}}).\ee

We make the following  expansion
for $\Phi$
\be
\Phi(t)=B_0+B _0 \sinh(rt)+B_2 \sinh(2rt)+......\ee
and solve (\ref{eq-3-j}) term by term. We have
\bea
&\,&e^{\tau\partial^2_t}_{\,\,\,0}\,(B_0+B _0 \sinh(rt)+B_2 \sinh(2rt)+...)\\
&=&-\mu_0-\mu_1{\mbox{erf}}(\frac{1}{2}\frac{t}{\sqrt{\tau}})+g(B_0+B _1 \sinh(rt)
+B_2 \sinh(2rt)+...)^2.
\eea
Comparing terms with erf and $\sinh(rt)$
we get the set of relations
\bea
B_0&=&\mu_1,\,\,\,\,\,\,\,\,\,\,\,\,\,\,\,\,\,\,
\mu_0=B_0-\frac{gB^2_1}{2}. \\
e^{\lambda r^2}B_1&=&2gB^2_0B _1,\,\,\,\,\,
e^{\lambda 4r^2}B_1=\frac{gB^2_1}{2}+2gB_0B_2,\eea
that gives
\bea
B_0&=&\mu_1,\,\,\,\,\,\,\,\,\,\,\,\,\,\,\,\,\,\,\,\,\,\,\,\,\,\,
B_1=\pm\sqrt{ \frac{2(\mu_0+\mu_1)}{g}},\\
r^2&=&\frac{\ln(2g\mu_1^2)}{\lambda },\,\,\,\,\,\,\,\,\,\,
B_2=
\frac{1}{2g\mu_1}\left(\mu_0+\mu_1
\mp(2g\mu_1^2)^4\sqrt{ \frac{2(\mu_0+\mu_1)}{g}}\right).\eea

\subsection{Neumann's case}
In this section we consider
\be
(-\xi^2\partial^2_t+m^2)e^{\lambda\partial^2_t}_{\,N\,J}\Phi=V^\prime(\Phi). \label{NSFT}
\ee
Taking into account (\ref{sol-n-der})
we get
\bea
(-\xi^2\partial^2_t+m^2)e^{\lambda\partial^2_t}_{\,N\,0}\Phi&=&V^\prime(\Phi)-J_{\xi}.
\label{NJ}\\
J_{\xi}&=&(-\xi^2\partial^2_t+m^2)J\eea
where
the source $J$ is given by (\ref{current-N}).
Assuming that $J$ admits an expansion
\be J_{NSFT}(t)=gj_0+gj _1
\cosh(\Omega t)+gj_2 \cosh(2\Omega t)+......\ee
(with some still unknown $\Omega$)
we make the similar  expansion for $\Phi$
\be \Phi(t)=B_0+B _1
\cosh(\Omega t)+B_2 \cosh(2\Omega t)+......\ee
and solve (\ref{eq-3-j}) term by
term. We get
\bea
\label{Omega-N}
\Omega^2&=&\frac{m^2}{\xi^2}+ (\frac{gj_1}{\xi^2B_1}+\frac{2g^2j_0}{\xi^2m^2}-\frac{g^2B^2_1}{\xi^2m^2})e^{\tau \Omega^2}
\\
B_0&=&-\frac{gj_0}{m^2}+\frac12\frac{gB^2_1}{m^2}\label{B0-m}
\\
B_2&=&\frac{gj_2}{4\xi^2\Omega^2-m^2}e^{-4\tau \Omega^2}\label{B2m}
\eea
and so on. Equation (\ref{Omega-N}) defines $\Omega^2$ perturbatively,
\be
\Omega^2=\frac{m^2}{\xi^2}+ \frac{gj_1}{\xi^2B_1}e^{\frac{\tau m^2}{\xi^2}}+...\ee
and the last equations define perturbatively coefficients. We can see that this perturbative series is similar to the local
perturbation series
for the equation
\be
\label{leq-3-j}
(-\xi^2\partial^2_t+m^2)\Phi=-gJ+g\Phi^2\ee

It is known that equation (\ref{leq-3-j}) exhibits very interesting chaotic behavior
for  special currents, that cannot be seen perturbatively.
We will mention about these properties in  Sect. 11.

\section {Nonlocal Equation on the Half-line in Friedmann Cosmology}
For simplicity we consider the action (\ref{string}) with $\xi=0$:
\be
\label{NLCM-d-m}
S=\frac{1}{g_4}\int
d^4x\sqrt{-g}\left\{m_p^2\frac{R}{2}+\frac{1}{\gamma}\left(\frac1{2}
\Phi e^{-\frac14\Box }\Phi-V(\Phi)\right)\right\}.
\ee
Here all coordinates are dimensionless, see notations in Sect.2,
$x \to  x/M_s$,
$
m_p^2=M_p^2/M_s^2
$, $\gamma$ is dimensionless coupling constant.

In the spatially flat Friedmann metric
\be
ds^2=-dt^2+a^2(t)dx^2
\ee
the dynamics in the model
 is described by a system of two
nonlinear nonlocal equations \cite{IA} for the
tachyon field
 and the Hubble parameter $H(t)=\dot{a}/a$
\begin{eqnarray}
\label{EOM_ST0approx_phi}
e^{-\frac{1}{4}{\cal D}}\Phi &=&V^\prime(\Phi),\,t>0
\\
\label{EOM_ST0approx}
3H^2&=&\frac{1}{\gamma m_p^2}\left(\frac{1}{2}\Phi e^{-\frac14\cal D }\Phi
+V(\Phi)+{\cal E}_1+{\cal E}_2\right),
\eea
where
\be{\cal D}=-\partial _t^2-3H(t)\partial_t,\,\,\,\,\,H=\frac{\partial_t a}{a}
\ee
and
\bea
\label{E1}
{\cal E}_{1}&=&  \frac{1}{8}\int_0^1 ds\left(\,
e^{\frac{s-2}{8}{\cal D}}
\Phi \,\right)\cdot
\left({\cal D}\,\,
e^{-\frac{1}{8} s {\cal D}} \Phi\right),\\
\label{E2}{\cal E}_2&=& \frac{1}{8} \int_0^1 ds\left(\partial_t
\,e^{\frac{s-2}{8}{\cal D}}
\Phi \right)\cdot
\left(\partial_t
e^{-\frac{1}{8}s{\cal D}} \Phi\right).
\end{eqnarray}
The nonlocal energy ${\cal E}_{1}$
plays the role of an extra potential term and ${\cal E}_{2}$ the role of the
kinetic term.

\subsection{Dirichlet's case}
Let us define the action of the operator $e^{-\tau\cal D}_{D,J}$ (compare with the definition of the same
operator on the whole axis used in \cite{Joukovskaya:2007nq,Barnaby:2008vs,MN:2008}) on the function $\varphi(t)$ as
\be
e^{-\tau\cal D}_{D,J}\varphi(t)=\Psi_{D,H}(t,\tau),
\ee
where $\Psi_{D,H}(t,\tau)$ is  a solution to
\be
\label{H-bc-ic-w-m}
\begin{cases}\frac{\partial}{\partial \tau}\Psi_{D,H}(t,\tau)=
-{\cal D}\Psi_{D,H}(t,\tau),\,\,\,\,\,t>0,\,\,\,\,\tau>0,\\
\Psi_{D,H}(t,0)=\varphi(t),\\
\Psi_{D,H}(0,\tau)=\mu(\tau).
\end{cases}\ee

Let us consider the case $H(t)=H_0=constant$.
Note that solutions to the first equations in (\ref{H-bc-ic-w-m})
and (\ref{HE-bc-ic}) are related via
\be
\Psi_{D,H_0}(t,\tau)=e^{-\frac32H_0t-\frac94H_0^2\tau}\Psi_{D}(t,\tau).\ee
Now we can write the solution to (\ref{H-bc-ic-w-m})
 for $H(t)=H_0$ in the form
\bea
\nonumber
\Psi_{D,H_0}(t,\tau)&=&e^{-\frac32H_0t-\frac94H_0^2\tau}\left[\frac{1}{\sqrt{4\pi \tau}}
\int _0^\infty \,
e^{\frac32H_0t^\prime}\varphi(t^\prime)\,
[e^{-\frac{(t-t')^2}{4\tau}}-e^{-\frac{(t+t')^2}{4\tau}}]dt'\right.\\
&+&\left.\frac{t}{\sqrt{4\pi }}\int _0^\tau\,
\frac{e^{\frac94H_0^2\tau'}\mu(\tau')}{(\tau-\tau')^{3/2}}e^{-\frac{t^2}{4(\tau-\tau')}}\,d\tau'\right].
\label{TBH0-sol}\eea
Denoting  $\Psi_{D,H_0}(t,\tau)=e^{-\tau{\cal D}}_{D,{\cal J}}\varphi(t)$
we can write
\be
e^{-\tau{\cal D}}_{N,{\cal J}} \varphi(t)=
e^{-\tau{\cal D}}_{N,0} \varphi(t)+{\cal J}(\tau,t),\ee
 where
\be
{\cal J}(\tau,t)= e^{-\frac32H_0t-\frac94H_0^2\tau}\frac{t}{\sqrt{4\pi }}\int _0^\tau\,
\frac{e^{\frac94H_0^2\tau'}\mu(\tau')}{(\tau-\tau')^{3/2}}\,e^{-\frac{t^2}{4(\tau-\tau')}}\,d\tau'.\ee
$$\,$$
The action of operator $e^{-\tau{\cal D}}_{D,0}$ on the function
$e^{-\frac32H_0t}\sinh(rt)$ is
given by the formula
\bea
e^{-\tau{\cal D}}_{D,0}e^{-\frac32H_0t}\sinh(rt)=
e^{\tau (r^2-\frac94H_0^2)}e^{-\frac32H_0t}\sinh (rt).
\label{sinh-mu-H0}
\eea

\subsection{Neumann's case}
In an analogous way  we
define the action of the Neunmann operator $e^{-\tau\cal D}_{N,\mu}$  on the function $\varphi(t)$ as
\be
e^{-\tau\cal D}_{N,\nu}\varphi(t)=\Psi _{N,H}(t,\tau),
\ee
where $\Psi_{N,H}(t,\tau)$ is  the solution to
\be
\label{H-bc-ic}
\begin{cases}\frac{\partial}{\partial \tau}\Psi_ {N,H}(t,\tau)=
-{\cal D}\Psi _{N,H}(t,\tau),\,\,\,\,\,t>0,\,\,\,\,\tau>0,\\
\Psi _{N,H}(t,0)=\varphi(t),\\
\frac{\partial}{\partial t}\Psi _{N,H}(t,\tau)|_{t=0}=\nu(\tau),
\end{cases}\ee
For $H(t)=H_0$  the solution of (\ref{H-bc-ic})
has the form
\bea
\nonumber
\Psi_ {N,H_0}(t,\tau)&=&e^{-3H_0t-\frac94H_0^2\tau}
\left[\frac{1}{\sqrt{4\pi \tau}}\int _0^\infty \,
e^{3H_0t^\prime}\varphi(t^\prime)\,
[e^{-\frac{(t-t')^2}{4\tau}}+e^{-\frac{(t+t')^2}{4\tau}}]dt'\right.\\
&+&\left.\frac{1}{\sqrt{\pi }}\int _0^\tau\,
\frac{e^{\frac94H_0^2\tau'}\nu(\tau')}{(\tau-\tau')^{1/2}}e^{-\frac{t^2}
{4(\tau-\tau')}}\,d\tau'\right].
\label{TBH0-sol}\eea
The action of the operator $e^{-\tau{\cal D}}_{N,0}$ on the function
$e^{-\frac32H_0t}\cosh(rt)$ is
given by
\bea
e^{-\tau{\cal D}}_{N,0}e^{-\frac32H_0t}\cosh(rt)=
e^{\tau (r^2-\frac94H_0^2)}e^{-\frac32H_0t}\cosh (rt).
\label{sinh-mu-H0}
\eea

\subsection {Solutions to Linearized Nonlocal Equation on the Half-line, $H_0\neq 0$}

Here we  consider the case of the Dirichlet
daemon with  $j(\mu|t,\tau)= 0$.
The equation of motion
\be\label{D0}
e^{\tau \cal D}_{D,0}\Phi_{H_0}(t)=m^2\Phi_{H_0}(t), \,\,t>0, \,\,\tau>0,
\ee
 where $\tau >0$ is a fixed parameter,
 has the form of the  following integral equation
\be
\label{N-H0}
\frac{e^{-\frac32H_0t-\frac94H_0^2\tau}}{2\sqrt{\tau\pi} } \int\limits_0^\infty
dt'\,e^{\frac32H_0t}[e^{-\frac{(t-t')^2}{4\tau}}-e^{-\frac{(t+t')^2}{4\tau}}] \,\Phi(t')
=m^2\Phi(t), \,\,\,\,t>0
\ee
From (\ref{sinh-mu-H0}) we see that
\be
\Phi_{H_0}(t)=B_1e^{-\frac32H_0t}\sinh(rt)
\ee
 solves equation
(\ref{N-H0}) if
\be
\label{r-H0}
(r^2-\frac94H_0^2)=\frac{\ln m^2}{\tau},
\ee
i.e. the spectrum of the operator $e^{\tau {\cal D}_{H_0}}_{D,0}$ is the same as
spectrum of the operator $e^{\tau {\cal D}_{H_0}}$ on the whole axis.

From this consideration we get that the energy and pressure
\bea
E_{D,H_0}(B_1)&=&-(r^2-\frac94H_0^2) e^{(r^2-\frac94H_0^2)}B^2_1,\\
P_{D,H_0}(B_1)&=&
-(r^2-\frac94H_0^2)B^2_{1}e^{(r^2-\frac94H_0^2)}\cosh
\left(2t\sqrt{r^2-\frac94H_0^2}\,\right).\eea
This energy and pressure can be used to find a deviation of the FRW metric from the dS one.

\section{Inflation and External Sources }

We have obtained in this  paper that in nonlocal string field theory
equations on the half-line a new arbitrary function appears which was called the
daemon external source. Let us now              discuss a possible role
of the external source in the cosmological inflation scenario.

The cosmological observations show that the universe is almost flat
and a density perturbations are scale invariant, Gaussian and adiabatic.
The simplest explanation of these observations is provided by the slow-roll
inflation driven by a scalar field \cite{inflation}.

There are two  versions of the inflation scenario with a single scalar field
\cite{Linde}-\cite{Rubakov}.
In the new inflation scenario one uses a large cosmological constant
(the simplest potential is $U(\phi)=U_0-m^2\phi^2/2$)
while the chaotic inflation scenario works with
the vanishing cosmological constant
(the simplest potential is $U(\phi)=m^2\phi^2/2$).

The basic ideas of the chaotic inflation scenario introduced
by Linde \cite{Linde-1,Linde-2} are very natural and general.
It is assumed that the early universe initially consisted of
many domains with chaotically distributed scalar field. Then
the rapid expansion (inflation) of the early universe
made from  the domains, in
which the scalar field was rather large,
many very big homogeneous domains.
Our observable universe is one of these big homogeneous domains.

A great advantage of the  chaotic inflation scenario
is its simplicity: it may occur
even in the theory of the free massive scalar field.
The  chaotic inflation
occurs in many models of the scalar field
with sufficiently flat potentials
and for a wide class of randomly chosen initial conditions.

It is supposed that the
natural initial conditions for the scalar field $\varphi_0$ at
the moment when the classical description of the universe first
becomes feasible is $U(\phi_0)\lesssim M_P^4$, where
$U(\phi)$ is the potential and $M_P$ is the Planck mass.
If $U(\phi)=g
\phi^n$, where  $n>1, 0<g<1$, then
it follows that the typical initial value $\phi_0$ of the field $\phi$ in the early
universe is  large, $\phi_0 >( M_P^4/g)^{1/n}$, greater than the Planck mass.

One could wish to get an inflation scenario with smaller
initial values of the field since at the Planck scale one doubts that the classical description is applicable.
We will  see that the external source can help to make smaller at least
one of the slow roll parameters $(\epsilon)$ even for
 rather small initial values of the scalar.

Given this picture of chaotically distributed scalar field
in the early universe we can ask ourself whether
other dynamical and external fields and  sources
could change the conclusions of the chaotic inflation scenario?
It seems that the basic result, that inflation
 created  our
modern homogeneous universe from a rather randomly
chosen domain in the initial chaos, still should be valid
in this more general situation  with arbitrary number
of chaotically distributed fields and sources.

Let us consider the action (\ref{L-l-inf-pairs}).
The Friedmann equations for
the homogeneous field $\phi=\phi(t)$ and the source $J=J(t)$ have the form
\bea
\ddot{\phi}&+&3H\dot{\phi}+U^{'}(\phi)-J=0,
 \label{Fr-eq}
\\
\label{H-F}
H^2&=&\frac{1}{3M_P^2}\left(\frac{\dot{\phi}^2}{2}
+U(\phi)-J\phi\right).
\eea
 The slow roll parameters are
\begin{equation}
\epsilon=\frac{M_P^2}{2}\left(\frac{U^{'}(\phi)-J}{U(\phi)-J\phi}\right)^2,\,\,\,
\eta=M_P^2
\frac{U^{''}(\phi)}{U(\phi)-J\phi}
 \label{Sl-roll}
\end{equation}
If the potential $U(\phi)=m^2\phi^2/2$ then
\be
\epsilon=
\frac{M_P^2}{2\phi^2}\frac{\delta^2}{(\frac{m^2}{2}\phi-\delta)^2}
 \label{Sl-roll-1}
\ee
where we denote $\delta=m^2\phi-J$. Therefore in the spacetime domain, where the daemon source
$J$ is such that  $\delta\ll m^2\phi/2$, we get a small parameter $\epsilon$
even for the not so
large field $\phi$. We can not make the parameter $\eta$ very small for the small
field $\phi$ by using a reasonable  choice
of the source $J$. To this end,  we still have to take the large values of the
field $\phi$ or the large cosmological constant. Note, that if the souse $J(t)$ in
(\ref{Fr-eq}) has the form
\be
J(t)=\sum _n\epsilon _n \cos(\omega _n t +\theta _n)\ee
and there is no  $J\phi$ term in (\ref{H-F}), one gets a nonlinear resonance \cite{Reichel}.
The external sources play the role of the perturbation in the FRW cosmology. Now, the result
of the action of this perturbation to the evolution of the Universe depends of the form of perturbation.
If the source behaves like a monotonic function then the daemon source can help to get
the chaotic inflation scenario with small scalar field. However, if the source is a periodic function, when
one could get the nonlinear resonance which destroys the inflation scenario.

Note also that if the current $J(x)$ in (\ref{Fr-eq}) is a white noise
with the correlation function
\begin{equation}
\langle J(x)J(y)\rangle=H^5\delta(x-y)
 \label{Sl-roll}
\end{equation}
and if one can neglect the term $J\phi$ in
then we obtain the Langevin equation in the stochastic approach to
inflation introduced by Starobinsky \cite{Starob}, for a general
discussion see \cite{Wood}. The stochastic approach fits well with
the functional approach to mechanics \cite{Vol-Funct}. A general
procedure of obtaining the quantum white noise from quantum field
theory is developed in the stochastic limit framework \cite{ALV}.

According to the anthropic principle perhaps there are many
universes and our universe is just one of them better suitable to
support life as we know it. Actually this is just a phenomenological
principle that the theory should be consistent with observations.
Various local fields $\psi_n$ in (\ref{l-inf-pairs}) accompanied by
the daemon sources could live in different universes developing
dynamics needed for the inflation scenario. If there is a
singularity also in future then it seems
 one has to set also a daemon boundary condition there.

\section{Conclusion}
 In this paper we have shown that string theory leads to nonlocal
 equations on the semi-axis which include an arbitrary source
 function $J(x)$.  At the moment we go not have a theoretical
 principle which could help to fix $J(x)$ except a phenomenological
 requirement to be consistent with the observed universe.

There are many interesting mathematical and physical questions to be considered
in the theory of the cosmological daemon. We hope to discuss them in future works.

\section{Acknowledgements}

We would  like to thank Branko Dragovich, Alexey Koshelev and Sergey Vernov  for the helpful discussions. The work is partially supported by grants RFFI 11-01-00894-a, NS 8265.2010.1 (I.A.) and RFFI 11-01-00828-a, NS 7675.2010.1 (I.V.).

\newpage
\appendix
\section{Laplace transformation and definition of $ F(\partial_t)$}
We remind that  the Laplace transform of the function  $\varphi(t)$, $t>0$
is defined as
\be
\label{L-tr}
\tilde \varphi(s)=\int_{0}^\infty\varphi(t)e^{-st}dt,\,\,\,\,{\mbox {Re}}s >a,
\ee
where $a$ is some number. The inverse transform is given by
\be
\label{lt-sym-r}
\varphi (t)=\frac{1}{2\pi i}\int_{c-i\infty}^{c+i\infty}
\tilde\varphi(s)e^{ts}ds,\ee
where $c>a$.

The action of derivatives is given by \cite{Evgraf}
\bea
\label{appder-r}
  \partial_t^{\, n}\phi(t) = \frac{1}{2\pi i}\int_{c-i\infty}^{c+i\infty} ds\, e^{st}
  s^n \left[ \tilde{\phi}(s) - \sum_{j=0}^{n-1 }\frac{\partial_t^{\,j}\phi(0) }{s^{j+1}}  \right]
\eea

Definition (\ref{appder-r}) has a meaning under the assumption that
the function
\be \Phi(z)=z^n\left[ \tilde{\phi}(z) - \sum_{j=0}^{n-1
}z^{-j-1}\partial_t^{\,j}\phi(0)  \right] \ee is holomorphic in $ Re
\,z >a$ and satisfies condition \be |\Phi(z)|={\cal
O}(|z|^{-1-\alpha}),\,\,\,for \,\,\,|z|\to\infty,\,\,\,a\leq Re\, z
\ee with $\alpha>0$ and $a\leq c$.

It follows from (\ref{appder-r}) that for an analytical function
\be
F(z)=\sum^\infty
 _{n=0}F_n\,z^n. \label{ser-F-r}
 \ee
it is natural to define the action of the operator $ F(\partial_t)$
as \cite{BK:2007}
\be
\label{ps-F}
  F(\partial_t)\,\varphi(t) = \frac{1}{2\pi  i}\int_{c-i\infty}^{c+i\infty}
   ds\, e^{st} \left\{ F(s)
  \tilde{\varphi}(s)-r(s )\right\},\ee
  where
  \be
  \label{ps-F-r}
  r(s )= \sum_{n=0}^{\infty} \sum_{j=1}^n
  F_n s^{n-j}\partial_t^{\,j-1}\varphi(0).
\ee

!!Applying  the general definition (\ref{ps-F}) to the case of exponential function
we obtain definition
(\ref{pseudoapp-exp}) presented in the text.!!
In \cite{Gorka} has been mentioned a more general definition of $ F(\partial_t)$
that also uses the representation (\ref{ps-F}) but the residual term
now is defined by infinite number of constants $d_j$
\be
  \label{ps-F-rd}
  r(s )= \sum_{n=0}^{\infty} \sum_{j=1}^n
  F_n d_{\,j-1}s^{n-j}.
\ee
For the case of the $\exp(\tau\partial^2_t)$
one gets the following definition
\be \label{pseudoapp-exp-G}
  e^{\tau\partial^2_t}_{\,\,\,\{LG\}}\varphi(t) =
  \frac{1}{2\pi i}\int\limits_{c-i\infty}^{c+i\infty} d\!s\, e^{st}\left[ e^{\tau s^2}  \tilde{\varphi}(s)
  -   \sum_{k=1}^{\infty} \sum_{j=1}^{2k}
  \frac{\tau^ks^{2k-j}}{k!}d_{j-1}  \right].
  \ee
Here $d_{j}$ are arbitrary constant and the subscript $LG$ indicates that the definition is based on the modification of formula
(\ref{pseudoapp-exp}) based on the Laplace transform.

\section{Proof of proposition 1.}
To prove  Proposition 1  we use the representation (\ref{pseudoapp-exp}) and the following two
lemmas.

{\bf Lemma 1.} One has
\be \label{exp}
\frac{1}{2\pi i}\int\limits_{c-i\infty}^{c+i\infty} d\!s\, e^{st} e^{\tau s^2}  \tilde{\varphi}(s)=
 \frac{1}{2\sqrt{\tau\pi} } \int\limits_0^\infty
dt'\,e^{-\frac{(t-t')^2}{4\tau}} \,\varphi(t'),\ee
{\it here } $\tilde{\varphi}(s)$   {\it is the Laplace transform of} $\varphi(t)$.

{\it Proof}. Substituting (\ref{L-tr}) to the LHS of (\ref{exp}) we get

\be \label{exp-m}
\frac{1}{2\pi i}\int\limits_{c-i\infty}^{c+i\infty} d\!s\, e^{st} e^{\tau s^2}  \tilde{\varphi}(s)
=\frac{1}{2\pi i}\int\limits_{c-i\infty}^{c+i\infty} d\!s\, e^{st} e^{\tau s^2}
\int_{0}^\infty\varphi(t^\prime)e^{-st^\prime}dt^\prime.
\ee
Assuming that we can interchange the order of integration
 and writing integral on $s$ as integral on $w$, $s=c+iw$ we can write
 \be \label{exp-mm}
\frac{1}{2\pi i}\int\limits_{c-i\infty}^{c+i\infty} d\!s\, e^{st} e^{\tau s^2}  \int_{0}^\infty\varphi(t^\prime)e^{-st^\prime}dt^\prime=
\frac{1}{2\pi }\int_{0}^\infty\int\limits_{-\infty}^{\infty} d\!s\, e^{(c+iw)(t-t^\prime)} e^{\tau (c+iw)^2}  \varphi(t^\prime)dt^\prime
\ee
$$=\frac{1}{2\sqrt{\pi \tau}}\int_{0}^\infty
e^{-\frac{(t-t^\prime)^2}{4\tau}}\varphi(t^\prime)dt^\prime.$$
That proves the Lemma 1.

{\bf Lemma 2.}
{\it  The function} $r(s,\tau)$
{\it which is  given by}
\be
\label{r-s}
 r(s,\tau)=\sum_{k=1}^{\infty} \sum_{j=1}^{2k}
  \frac{\tau^ks^{2k-j}}{k!}\varphi^{(j-1)}(0)
  \ee
{\it admits the following representation}
\be
 \label{I-rr}
  r(s,\tau)=
\int _0^\tau \,d\tau'e^{s^2(\tau-\tau')}(\mathfrak{j}_0(\tau')
+ s\mathfrak{j}_1(\tau')),
 \ee
{\it where}
\bea
  \label{j-0}
  \mathfrak{j}_0(\tau)&=&\sum _{k=0}^{\infty}
\frac{\tau^{k}}{k!}\,\lim_{t\to +0}\partial _t^{2k+1}\varphi(t),\\
\label{j-1}
  \mathfrak{j}_1(\tau)&=&
\sum _{k=0}^{\infty} \frac{y^{k}}{k!} \, \lim_{t\to +0} \partial _t^{2k}\varphi(t).
\eea

{\it Proof.} Formula (\ref{r-s}) can be rewritten in the following way
\bea
\nonumber
  r(s,\tau)&=&\sum_{k=0}^{\infty}\left[
  \varphi ^{(2k)}(0)  \frac{e^{s^2\tau }-1-s^2\tau-\frac{1}{2!}(s^2\tau)^2-...-\frac{1}{k!}(s^2\tau)^k}{s^{2k+1}}
  \right.\\
  \label{r-exp-3}
\,\,\,\,\,\,\,\,&\,& \left.+
\varphi^{(2k+1)}(0)  \frac{e^{s^2\tau }-1-s^2\tau-\frac{1}{2!}(s^2\tau)^2-...-\frac{1}{k!}(s^2\tau)^k}{s^{2k+2}}
  \right].\eea
Using
 \be
  e^{s^2\tau}\int _0^\tau e^{-s^2y}\frac{y^{m}}{m!}\,dy=\frac{e^{s^2\tau}-1-s^2\tau-\frac{1}{2}(s^2\tau)^2-\frac{1}{3!}(s^2\tau)^3
  ...\,-\frac{1}{m!}(s^2\tau)^{m}}{s^{2m+2}}
\ee
we can rewrite $r(s,\tau)$ as
\bea
\nonumber
r(s,\tau)&=&\sum _{k=0}^{\infty}e^{s^2\tau}\partial _t^{2k}\varphi(0)\,s\,
\int _0^\tau e^{-s^2y}\frac{y^{k}}{k!}\,dy\\
\label{r-exp-4}&+&
\sum _{k=0}^{\infty}    e^{s^2\tau}\partial _t^{2k+1}\varphi(0)
\,\int _0^\tau e^{-s^2y}\frac{y^{k}}{k!}\,dy\eea
that gives the proof of Lemma 2.

{\bf Lemma 3.} {\it The functions }$\mathfrak{j}_0$ {\it and }$\mathfrak{j}_1,$
{\it given by (\ref{j-0}) and (\ref{j-1}), admit the following representation}
\bea
\label{j0}
\mathfrak{j}_0(\tau')=\frac{\phi (\sqrt{\tau' })-
\phi (-\sqrt{\tau' })}{2\sqrt{\tau' }},\,\,\,
\mathfrak{j}_1(\tau')=\frac{\phi (\sqrt{\tau' })+
 \phi (-\sqrt{\tau' })}{2},
 \eea
{\it where $\phi$ is related to $\varphi$ by (\ref{phi-varphi})}.

{\it Proof.} Substituting into (\ref{r-exp-4}) the relations
\be
\label{univ-phi}\phi_{n}=\frac{n!}{[\frac{n}{2}]!}\varphi_{n}
\ee
we get the representation in terms of $\phi$
\bea
\nonumber
  r(s,\tau)&=&\sum _{k=0}e^{s^2\tau}\phi_{2k}\,
\int _0^\tau e^{-s^2y}\frac{s^{k}y^{2k}}{(2k)!}\,dy\\
\label{r-exp-5}&+&
\sum _{k=0}    e^{s^2\tau}\phi_{2k+1}
\,\int _0^\tau e^{-s^2y}\frac{s^{k}y^{2k+1}}{(2k+1)!}\,dy.\eea
Taking into account that
\be\sum _{k=0}^\infty \phi_{2k+1}\frac{\tau ^k}{(2k+1)!}=\frac{\phi (\sqrt{\tau })
- \phi (-\sqrt{\tau })}{2\sqrt{\tau }}
\ee
and
\be\sum _{k=0}^\infty \phi_{2k}\frac{\tau ^{\frac{2k}{2}}}
{(2k)!}=\frac{\phi (\sqrt{\tau })
+ \phi (-\sqrt{\tau })}{2}
\ee
we get
 \be
 \label{I-rr}
  r(s,\tau)=
\int _0^\tau \,d\tau'e^{s^2(\tau-\tau')}
\left[\frac{\phi (\sqrt{\tau' })
- \phi (-\sqrt{\tau' })}{2\sqrt{\tau' }}+ s
\frac{\phi (\sqrt{\tau' })+
 \phi (-\sqrt{\tau' })}{2}\right],
 \ee
that proves Lemma 3.

{\bf Lemma 4.}

{\it The second term  in the integrand in the RHS of}
(\ref{pseudoapp-exp})
\be
 r_L(t,\tau)=\frac{1}{2\pi i}\int\limits_{c-i\infty}^{c+i\infty}
 d\!s\, e^{st} r(s,\tau)
\ee
  {\it admits the following integral representation}
\bea
\label{r-x}
  r_L(t,\tau)=\frac{1}{2\sqrt{\pi }}\int _0^\tau \,d\tau'
e^{-\frac14\frac{t^2}{(\tau-\tau')}} \frac{\mathfrak{j}_0(\tau')}
{\sqrt{\tau-\tau'}}
-\frac{t}{4\sqrt{\pi }}\int _0^\tau \,d\tau'
e^{-\frac14\frac{t^2}{(\tau-\tau')}}
\frac{\mathfrak{j}_1(\tau')}{(\tau-\tau')^{3/2}}.
\eea

The proof follows from the representation (\ref{I-rr}) by computing the Laplace
transform.
Indeed we have
\bea
\nonumber
  r_L(t,\tau)&=&\frac{1}{2\pi i}\int_{c-i\infty}^{c+i\infty} ds\, e^{st}r(s,\tau)
  \\
\nonumber&=&\frac{1}{2\pi i}\int_{c-i\infty}^{c+i\infty} ds\, e^{st}
\int _0^\tau \,d\tau'e^{s^2(\tau-\tau')} \mathfrak{j}(\tau',s)\\\nonumber
&=&\frac{1}{2\pi i}\int_{c-i\infty}^{c+i\infty} ds\, e^{st}
\int _0^\tau \,d\tau'e^{s^2(\tau-\tau')} [\mathfrak{j}_0
(\tau')+s\mathfrak{j}_1(\tau')]\\
&=&\frac{1}{2\sqrt{\pi }}\int _0^\tau \,d\tau'
e^{-\frac14\frac{t^2}{(\tau-\tau')}} \frac{\mathfrak{j}_0(\tau')}{\sqrt{\tau-\tau'}}
-\frac{t}{4\sqrt{\pi }}\int _0^\tau \,d\tau'
e^{-\frac14\frac{t^2}{(\tau-\tau')}}
\frac{\mathfrak{j}_1(\tau')}{(\tau-\tau')^{3/2}}.\nonumber\eea
Therefore, we get
\bea
\label{r-x}
  r_L(t,\tau)=\frac{1}{2\sqrt{\pi }}\int _0^\tau \,d\tau'
e^{-\frac14\frac{t^2}{(\tau-\tau')}} \frac{\mathfrak{j}_0(\tau')}
{\sqrt{\tau-\tau'}}
-\frac{t}{4\sqrt{\pi }}\int _0^\tau \,d\tau'
e^{-\frac14\frac{t^2}{(\tau-\tau')}}
\frac{\mathfrak{j}_1(\tau')}{(\tau-\tau')^{3/2}}.
\eea
Lemmas 1,2,3 lead to the following representation
 \bea
\label{act-m}
e^{\tau\partial^2_t}_{\,\,\,L}\varphi(t) &=&\frac{1}{2\sqrt{\tau\pi} }
\int_0^\infty dt^\prime\,e^{-\frac{(t-t^\prime)^2}{4\tau}} \,\varphi(t^\prime)\\
\nonumber
&-&\frac{1}{4\sqrt{\pi }}\int _0^\tau \,d\tau'
e^{-\frac14\frac{t^2}{(\tau-\tau')}}
\frac{\phi (\sqrt{\tau' })-\phi (-\sqrt{\tau' })}
{\sqrt{\tau' }\sqrt{\tau-\tau'}}
\\
\nonumber
&+&\frac{t}{8\sqrt{\pi }}\int _0^\tau \,d\tau'
e^{-\frac14\frac{t^2}{(\tau-\tau')}}
\frac{\phi (\sqrt{\tau' })+
 \phi (-\sqrt{\tau' })}{(\tau-\tau')^{3/2}}.
\eea

{\it Proof of Proposition 1. }
We use   the fact, that the inverse Laplace transform gives $0$ for $t<0$
\bea
\label{ad-rel}
 0&=&\frac{1}{2\sqrt{\tau\pi} }
\int_0^\infty dt^\prime\,e^{-\frac{(t+t^\prime)^2}{4\tau}} \,\varphi(t^\prime)\\
&-&\frac{1}{4\sqrt{\pi }}\int _0^\tau \,d\tau'
e^{-\frac14\frac{t^2}{(\tau-\tau')}}
\frac{\phi (\sqrt{\tau' })- \phi (-\sqrt{\tau' })}{\sqrt{\tau' }\sqrt{\tau-\tau'}}
\\&-&\frac{t}{8\sqrt{\pi }}\int _0^\tau \,d\tau'
e^{-\frac14\frac{t^2}{(\tau-\tau')}}
\frac{\phi (\sqrt{\tau' })+
 \phi (-\sqrt{\tau' })}{(\tau-\tau')^{3/2}}.\eea
 Therefore, subtracting (\ref{ad-rel}) from (\ref{act-m})
 we get
  \bea
\label{Act}
\Psi_{N}(t,\tau)=e^{\tau\partial^2_t}\varphi(t) &=&\frac{1}{2\sqrt{\tau\pi} }
\int_0^\infty dt^\prime\,e^{-\frac{(t-t^\prime)^2}{4\tau}} \,\varphi(t^\prime)-
\frac{1}{2\sqrt{\tau\pi} }
\int_0^\infty dt^\prime\,e^{-\frac{(t+t^\prime)^2}{4\tau}} \,\varphi(t^\prime)\\
\nonumber
\\&+&\frac{t}{4\sqrt{\pi }}\int _0^\tau \,d\tau'
e^{-\frac14\frac{t^2}{(\tau-\tau')}}
\frac{\phi (\sqrt{\tau' })+
 \phi (-\sqrt{\tau' })}{(\tau-\tau')^{3/2}}.
\nonumber\eea
that completes the proof of  Proposition 1.

{\bf Lemma 5.} {\it The action of operator $e^{\tau\partial^2_t}_{\,\,\,L}$
 on the constant is the same constant,}
\be
e^{\tau\partial^2_t}_{\,\,\,L}1=1.\ee

{\it Proof.}
Taking into account that
\bea
\label{Act-11}
\frac{1}{2\sqrt{\tau\pi} } \int\limits_0^\infty
dt'\,[e^{-\frac{(t-t')^2}{4\tau}}-e^{-\frac{(t+t')^2}{4\tau}}]
&=&
{\mbox{erf}}(\frac{1}{2}\frac{t}{\sqrt{\tau}}),
\\
\label{Act-12}
\frac{t}{2\sqrt{\pi }}\int\limits_0^\tau \,d\tau'
e^{-\frac14\frac{t^2}{(\tau-\tau')}}\, \frac{1}{(\tau-\tau')^{3/2}}
&=&-{\mbox{erf}}(\frac{1}{2}\frac{t}{\sqrt{\tau}})+1\eea
we get
\be
\label{ActL1} e^{\tau\partial^2_t}_{\,\,\,L}1=
{\mbox{erf}}(\frac{1}{2}\frac{t}{\sqrt{\tau}})-
{\mbox{erf}}(\frac{1}{2}\frac{t}{\sqrt{\tau}})+1=1.
\ee
Here ${\mbox{erf}}(t)$ is
 the error function,
\be
\label{erf}
{\mbox{erf}}(t)=\frac{2}{\sqrt{\pi}}\int_0^te^{-x^2}dx.
\ee
Note that  ${\mbox{erf}}(-t)=-{\mbox{erf}}(t)$.

\end{document}